\documentclass[aip,aps,preprint,showpacs]{revtex4}
\usepackage{amssymb}

\usepackage{graphicx}

\usepackage[T1]{fontenc}

\usepackage[american]{babel}

\begin{document}

\title{Microstructure of epitaxial strained BiCrO$_3$ thin films}

\author{A.~David}
\affiliation{Laboratoire CRISMAT, CNRS UMR 6508, ENSICAEN, 6 Bd Mar$\acute{e}$chal Juin, F-14050 Caen Cedex 4.} 
\author{Ph.~Boullay}\thanks{philippe.boullay@ensicaen.fr} 
\affiliation{Laboratoire CRISMAT, CNRS UMR 6508, ENSICAEN, 6 Bd Mar$\acute{e}$chal Juin, F-14050 Caen Cedex 4.} 
\author{R.V.K.~Mangalam}
\affiliation{Laboratoire CRISMAT, CNRS UMR 6508, ENSICAEN, 6 Bd Mar$\acute{e}$chal Juin, F-14050 Caen Cedex 4.}
\author{N.~Barrier}
\affiliation{Laboratoire CRISMAT, CNRS UMR 6508, ENSICAEN, 6 Bd Mar$\acute{e}$chal Juin, F-14050 Caen Cedex 4.} 
\author{W.~Prellier}\thanks{wilfrid.prellier@ensicaen.fr} 
\affiliation{Laboratoire CRISMAT, CNRS UMR 6508, ENSICAEN, 6 Bd Mar$\acute{e}$chal Juin, F-14050 Caen Cedex 4.}

\date{\today}

\begin{abstract}
\quad The structure and microstructure of fully-strained BiCrO$_3$ thin films have been investigated by X-rays diffraction and
 transmission electron diffraction, at room temperature. 
Interestingly, three structural variants are simultaneously stabilized within the film. While two of them are consistent with the existing phases in the bulk-below and above the 420 K structural transition-a different 
phase is identified. The existence of various structures has been attributed to the inhomogeneous distribution of local strains and oxygens resulting from a minimization of 
the strain-energy at the interface. These findings will open the route to a better understanding of Bi-based perovskites and metastable phases.


\end{abstract}


\maketitle

\bigskip



Functional materials are of interest owing to their potential device applications, including non-volatile memories, magnetic read heads etc.
Among them, multiferroics - a material exhibiting both ferroelectricity and magnetism - have recently attracted some interest.
However, ferroelectricity and magnetism hardly coexist due to the structural competition in perovskites. 
Ferroelectricity needs a non-centrosymmetric structural distortion usually not compatible with the presence of partially filled d-levels ions required for magnetic ordering.\cite{spaldin}
Even known for more than forty years, the magnetic transition metal perovskites BiMO$_3$ (M=Mn, Fe, Cr, ...) have attracted much attention during the last decade as potential multiferroic materials. 
In these compounds, the presence of the stereochemically active electron lone pair $6s^2$ of Bi$^{3+}$ ions induce structural distortions likely to promote a non-centrosymmetric structure compatible with the existence of ferroelectricity. To date only BiFeO$_3$ has been clearly identified as a ferroelectric with antiferromagnetic ordering \cite{teague, wang, lebeugle} while the case of BiMnO$_3$ is still the subject of much debate.\cite{hill, santos, montanari, yang}

BiCrO$_3$ is another compound for which few studies have been reported.\cite{Cr-first} Also, the existence of a non-centrosymmetric structure together with a magnetic ordering of the Cr$^{3+}$ ion was unclear until the recent works of Belik et al.\cite{belik1, belik3} who established that BiCrO$_3$ exhibits a first order structural phase transition near 420 K 
with a step in the inverse magnetic susceptibility curves within the paramagnetic domain.\cite{Cr-first} 
Below this temperature, BiCrO$_3$ adopts a centrosymmetric $C2/c$ monoclinic structure (a = 9.464 Å, b = 5.479 Å, c = 9.585 Å and $\beta$= 108.568$^\circ$) while above a centrosymmetric $Pnma$ orthorhombic structure (a = 5.545 Å, b = 7.757 Å, c = 5.428 Å) is observed.\cite{belik3}  
The broad anomalies observed in the dielectric properties close to this transition \cite{niitaka} has been attributed by Belik et al. to a paraelectric-antiferroelectric phase transition.\cite{belik3} 
It is also established that an antiferromagnetic transition (T$_N=$) appears in the range of 109-123 K \cite{Cr-first, niitaka, belik1} followed by more complex magnetic transitions at lower temperatures.
 Below the Neel temperature, BiCrO$_3$ adopts a G-type antiferromagnetic structure with the magnetic moment of Cr$^{3+}$ aligned along the monoclinic $b$ axis with a weak canting angle.\cite{belik3}  

In bulk, BiCrO$_3$ is prepared using high-pressure technique (4-6 GPa). Such metastable phases can also be obtained in thin films using substrate-induced strains as shown in the case of BiMnO$_3$ \cite{Mn-film} and many other examples in cuprates, manganites or cobalites.\cite{wil,clara} Thus, thin films of BiCrO$_3$ were successfully grown using the non-equilibrium process of the laser ablation on (100)-LaAlO$_3$ with or without a (La$_{0.5}$Sr$_{0.5}$)CoO$_3$ buffer layer by Murakami et al. \cite{murakami} and on (100)-SrTiO$_3$ with a SrRuO$_3$ buffer layer by Kim et al.\cite{kim} 
In these works, the optimal growth conditions are 650$^\circ$C with an oxygen pressure about 1 to 5 mTorr.
In comparison to the bulk data, the epitaxial BiCrO$_3$ films shows a small increase at the magnetic transition temperature at 120K \cite{murakami} or 140 K \cite{kim} with a weak magnetic moment of the film consistent with the picture of a canted antiferromagnetic behaviour.
In reference \cite{murakami}, Murakami et al. reported using piezoelectric force microscopy a ferroelectric behavior, while Kim et al. in \cite{kim} found a double hysteresis loop in both the electric-field dependence of the dielectric constant, {\Large $\varepsilon_r$}($\it{E}$), and the polarization, $\it{P(E)}$, consistent with an antiferroelectric behaviour. 
Both authors claimed a triclinic unit cell for BiCrO$_3$ films despite any deep analysis. Nevertheless, these results are in agreement with the bulk structure as reported in the work of Sugawara et al. at the end of the sixties\cite{Cr-first} but different than the recent work of Belik et al.\cite{belik3}. 
The need of deeper structural analysis on BiCrO$_ 3$ films, which is a prerequirement to understand the physical properties of such films, has motivated our work.
In the present paper, the detailed microstructural study of high quality epitaxial BiCrO$_3$ thin films is presented using Transmission Electron Microscopy (TEM). The coexistence at room temperature, of the two structural forms (monoclinic and orthorhombic) is evidenced, in agreement with the bulk measurements of Belik et al.. More interesting, another phase is observed whose structure has not been reported so far in bulk. 
We explained our results using the substrate-induced strain that leads to an inhomogeneous distribution of the local strains and the oxygen vacancies.


200nm thin films of BiCrO$_3$ (BCO) were grown using the pulsed laser deposition technique (PLD) onto (100)-oriented SrTiO$_3$ (STO) substrates. 
Due to the volatility of the bismuth element, a sintered target with 5\% excess of bismuth were used.\cite{kim, ranjith} 
The preliminary structural study was performed by X-ray diffraction (XRD) using a Seifert XRD3000 diffractometer (Cu K$\alpha_1$ radiation) for the $\theta$-2$\theta$ measurements. 
The XRD patterns reveal the epitaxial growth of the films onto the (100)-oriented STO substrates
with a high degree of texture and a good crystalline quality as confirmed by the full-width-at-half-maximum around 0.12 close to the substrate value (0.1). 
The out-of-plane lattice parameter was determined to be equal to 3.88 Å very close to the substrate value, in agreement with those reported in the literature,\cite {murakami,kim} suggesting a strained film.



\begin{figure}
\includegraphics[angle=0,width=0.48\textwidth]{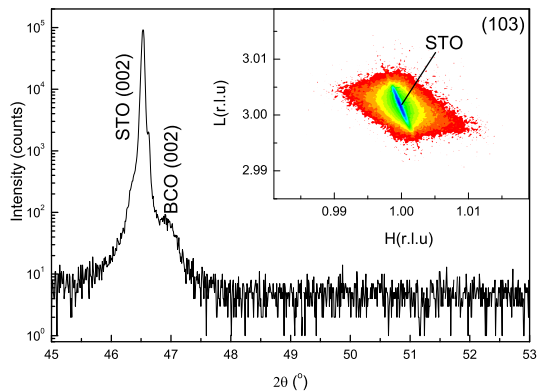}
\caption{\label{fig:space_map} XRD $\theta-2\theta$ showing that the out-of-plane parameter is close to substrate. Insert: reciprocal space map recorded around (103) reflection as an isointensity contour plot in logarithmic scale for a 200 nm sample. The logarithmic scale ranges from 3 to 4000. The Laue indices H and L are defined the lattice parameters of the STO substrate.}
\end{figure}
 



The $\theta$-2$\theta$ scan (Figure~\ref{fig:space_map}) shows the epitaxial growth of BCO thin-film (00\textit{l}) without detectable impurity phase.
Diffraction peaks are very in close to the STO substrate.The azimuthal $\phi$-scan recorded along (103) reflections reveals a fourfold symmetry demonstrating the in-plane alignement, cube-on-cube growth
of the BCO unit-cells and among the film and the STO substrate in-plane crystallographic axes (not shown). 
The reciprocal space map along the substrate (103) reflection (insert of fig.~\ref{fig:space_map}) reveals that the BiCrO$_3$ film has grown 
in close match with the in-plane and out-of-plane lattice parameters of STO substrate. This confirms that the BiCrO$_3$ film is fully-strained in comparison to bulk lattice parameters.


\begin{figure}
\includegraphics[width=0.48\textwidth]{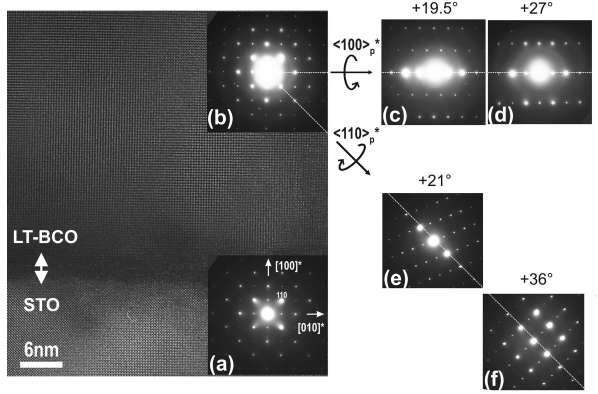}
\caption{\label{fig:BCOLT-TEM} HREM image of the STO/LT-BCO interface observed along [001]$_\mathrm{STO}$ with the corresponding SAED patterns of the substrate and the film in (a) and (b), respectively. On the right of the picture, the patterns (c-d) and (e-f) obtained by tilting the film around a direction <100>$^*_\mathrm{p}$ and <110>$^*_\mathrm{p}$, respectively, 
can be indexed considering the LT-BCO monoclinic form of BiCrO$_3$ \cite{belik3} as: (c) [141] or [21$\bar{1}$], (d) [161] or [32$\bar{1}$], (e) [152] or [$\bar{1}$11] and (f) [031] or [$\bar{1}$21] considering (b) compatible with both zone axes [121] or [10$\bar{1}$].}
\end{figure}


In order to obtain a better insight on the nature of the deposited phases and their epitaxial relation, 
cross-sectional samples from different films were further investigated by TEM (FEI Tecnai G2 30 and Jeol 2010) using
Selected Area Electron Diffraction (SAED) and High Resolution Electron Microscopy (HREM).
The high quality of the as-deposited films is confirmed by the absence of 
secondary phase such as Bi$_2$O$_3$ at the interface or within the film, and by the well-defined, sharp and coherent interface. The EDS analyses showed that the ratio Bi/Cr is largely homogeneous and equal to 1. 
Nonetheless, both SAED and HREM observation reveal the presence of three structurally distinct phases: the low temperature phase, noted LT-BCO, the high temperature phase,
noted HT-BCO and a third phase, noted X-BCO, which structure is discussed hereafter.

\begin{figure}
\includegraphics[width=0.48\textwidth]{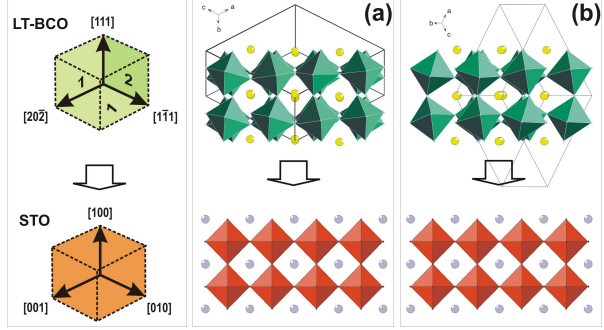}
\caption{\label{fig:BCOLT-epitaxie} (color online) In the left part, schematic representation describing the epitaxial relation between LT-BCO and STO as deduced from the TEM observations. The two possibilities, denoted (a) and (b), envisaged for an epitaxial growth of LT-BCO onto STO differ by having the direction $\sim$[$\bar{1}$01]LT-BCO lying in-plane or out-of-plane, respectively.}
\end{figure}

Figure~\ref{fig:BCOLT-TEM} displays a typical HREM image showing an interface between STO and LT-BCO for a <100>$_\mathrm{STO}$ zone axis orientation. The corresponding SAED patterns are given in the insert.
 Both SAED patterns look similar with in-plane and out-plane lattice parameters in the film close to 3.9Å, which indicates a pseudo ``cube-on-cube'' growth of this phase onto the STO substrate. 
In order to determine its crystal structure, a reconstruction of the reciprocal space was done by tilting the sample at specific angles. The resulting series of SAED patterns (Fig.~\ref{fig:BCOLT-TEM}) and the comparison with simulations (not shown) reveals that actually LT-BCO possess a monoclinic unit cell compatible with the low temperature phase reported in bulk BiCrO$_3$.\cite{belik3}
The SAED patterns of the film (Fig.~\ref{fig:BCOLT-TEM}b) recorded along the [001]$_\mathrm{STO}$ zone axis is indeed compatible with either the [121] and the [10$\bar{1}$] zone axes patterns of the BiCrO$_3$ monoclinic form.  
Assuming that each orientation is possible(see schematic drawing in Fig.~\ref{fig:BCOLT-epitaxie}), the epitaxial relationship for LT-BCO can be described as follow:
 first (111)LT-BCO$\parallel$ (100)STO with $\sim$[$\bar{1}$01]LT-BCO$\parallel$ [010]STO
and second (10$\bar{1}$)LT-BCO$\parallel$ (100)STO with $\sim$[121]LT-BCO$\parallel$ [010]STO. This epitaxial relationships are similar to those reported for the epitaxial growth of BiMnO$_3$ thin films on (100)-STO.\cite{dossantos2}
The calculation of mismatch, with respect to the bulk 
between STO and the LT-BCO gives -0.6\% and -0.2\%, respectively. Since these values are very close, both orientation variants of this form are expected to be present in the film.


\begin{figure}
\includegraphics[width=0.47\textwidth]{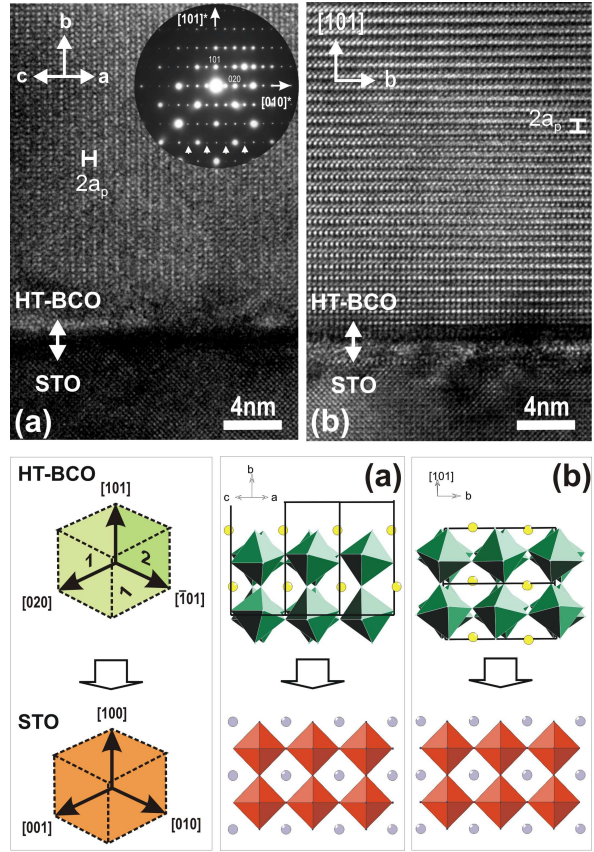}
\caption{\label{fig:BCOHT-TEM} (color online) Upper part: HREM images illustrating the epitaxial growth of HT-BCO on the STO substrate. The two possibilities, in (a) and (b), differ by having the direction [010] HT-BCO lying out-of-plane or in-plane, respectively. In (a), the insert represents the SAED patterns corresponding to the HREM image. Rows of additional reflections as compared to a <100>$_\mathrm{STO}$ zone axis orientation are indicated by white arrows. Lower part: schematic representation describing the epitaxial relation between HT-BCO and STO as deduced from the TEM observations.}
\end{figure}

For the second form (denoted HT-BCO) both HREM image and SAED patterns obtained for a <100>$_\mathrm{STO}$ zone axis orientation clearly indicate that the film possess a different structure as compare to the STO substrate (see Fig.~\ref{fig:BCOHT-TEM}).   
On the SAED patterns in the insert Fig.~\ref{fig:BCOHT-TEM}-1, rows of additional reflections are observed corresponding to the doubling of the in-plane lattice parameter (calculated to be equal to 2 $\times$ a$_p$ $\approx$ 7.8 Å). In Fig.~\ref{fig:BCOHT-TEM}-2, this is the out-of-plane lattice parameter that corresponds to $\approx$ 7.8 Å.
Using reciprocal space reconstruction by tilting experiments (not shown), we found that this part of the film is compatible with the high temperature form of BiCrO$_3$ reported in the bulk \cite{belik3} and having an orthorhombic unit cell as described above.  
Considering each possible orientation (see schematic drawing in the left part of Fig.~\ref{fig:BCOHT-TEM}), the epitaxial relationship for HT-BCO can be described as :
 first (101)HT-BCO $\parallel$ (100)STO with [010]HT-BCO $\parallel$ [010]STO
and second (010)HT-BCO $\parallel$ (100)STO with [101]HT-BCO $\parallel$ [010]STO. The calculation of the in-plane lattice parameter mismatch between STO and the HT-BCO gives -0.8\% in both cases.


\begin{figure}
\includegraphics[width=0.48\textwidth]{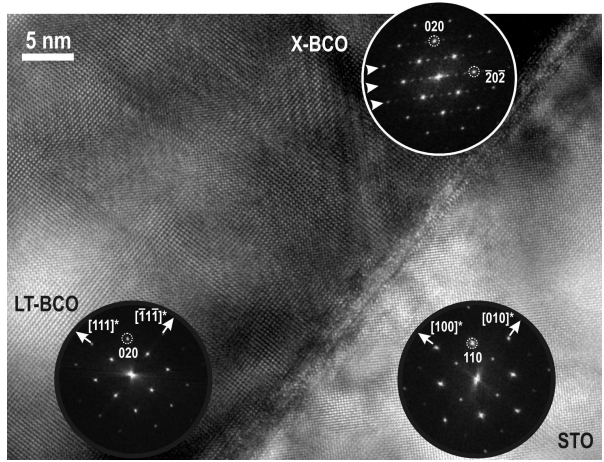}
\caption{\label{fig:BCOX-TEM} [001]$_\mathrm{STO}$ HREM image showing the interfaces BCO/STO and LT-BCO/X-BCO, which are both parallel to \{100\} STO planes. The Fourier Transform correspond to the diffraction patterns obtain from three areas taken in STO, LT-BCO and X-BCO. The main spots of the X-BCO form can be indexed considering the LT-BCO monoclinic cell plus additional satellites reflections.}
\end{figure}

The Figure~\ref{fig:BCOX-TEM} displays a HREM image revealing the presence (see in the upper part of the image) of a third BCO form (X-BCO) having a HREM contrast different from that of LT-BCO (this later phase is visible in the lower left corner of the figure). Less commonly observed, this form exhibits a strong structural relationship with LT-BCO as seen from the coherent interface existing between theses two forms. 
Actually, reciprocal space reconstruction shows that the strongest diffraction spots can be indexed considering the same $C2/c$ monoclinic cell. Nonetheless the Fourier Transform (FT) shows that X-BCO exhibits extra spots along the [101]* LT-BCO direction (see white arrows in the X-BCO FT Fig.~\ref{fig:BCOX-TEM}) that multiply d$_\mathrm{202}$ by four. Similar feature is observed on most of the reciprocal planes with main spots compatible with the monoclinic LT-BCO form in addition to satellites spots. 
To our knowledge, a structure related to BiCrO$_3$ compatible with the presence of these extra spots has not been reported yet.
 Our best guess is to consider a commensurate modulated structure related to LT-BCO and having a C2/c($\alpha$0$\gamma$)00 super space group with cell parameter similar to LT-BCO. Assuming a modulation vector in the form q=$\alpha$a*+$\gamma$c*, the determination of the $\alpha$ and $\gamma$ value is not trivial. Hence the extra spots in the FT Fig.~\ref{fig:BCOX-TEM} was indexed by taking $\alpha$=$\gamma$=1/2, which is the most straightforward choice.
 But other possibilities such as $\alpha$=1 and $\gamma$=0 would lead to the same patterns for this [10$\bar{1}$] zone axis. However, with the tilting experiments performed by SAED, a unique  solution could not be reached. 
This point has to be further investigated. It is intereting to note that this structure is closely related to the monoclinic form 
observed recently in BiMnO$_3$ at room temperature by high pressure synchrotron studies. \cite{belik4} For more decription,
BiMnO$_3$ retains in monoclinic $C2/c$ phase at ambient pressure, a $P$-phase {\it (P2$_1$/c)} appears above 1.5 GPa and an 
other transition occurs to $Pnma$ SG above 5.5 GPa. Similary, different phases of BiCrO$_3$ are observed in our strained films at room temperature. 
Assuming that each of phase is stabilized under a given pressure (or a given strains), an inhomogeneous strains distibution would not be sufficient to explain these results. 
Oxygen distributions will also play a key role which is most probably associated to the strains. 
In other words, the stabilization of mestastable phase is a result of strains and oxygen vacancies, both effect leading to a minimization of the
interface energy.


\quad In summary, well-crystallized epitaxial BiCrO$_3$ thin film have been deposited on (100)-oriented SrTiO$_3$ substrate 
by the pulsed laser deposition technique.
The microstructural study reveals the stabilization of the different structural phases within the fully strained film matrix:
a monoclinic $C2/c$ structure and an orthorhombic $Pnma$ structure, compatible with the high and low temperature phases,
respectively, found in the bulk compound. We also evidenced a third phase similar to a superstructure which is closely related
to the monoclnic $C2/c$ form. We hope that our findings will emphasis the route to
a better understanding in the multiferroic Bi-based perovskite phases.\\

We thank L. Gouleuf and J. Lecourt for technical support. This work is carried out in the frame of the STREP MACOMUFI (NMP4-CT-2006-313321) supported by the European Commission and by the CNRS, France. Financial support from GDR Multiferroiques (3163), C'Nano Nord-Ouest (2975), SONDE (ANR-06-BLAN-0331), CEFIPRA/IFPCAR (3908-1) and STAR (21465YL) projects are also acknowledged.

\end{document}